\documentstyle{l-aa}
\def\cm3{\,{\rm cm^{-3}}}

 \def\C34S{\,{\rm C^{34}S}} 
\def\13CS{\,{\rm ^{13}CS}} 
 
\def\sfir{\,{\it \sigma _{\rm FIR}}}
\def\td{\,{\it T_{\rm d}}}
\def\tdo{\,{\it T_{\rm do}}}
\def\h2{\,{\rm H_{2}}}
\def\nH2{\,{\it N(H2)}}
\def\nh{\,{\it N_{\rm H}}}
\def\wmsr{\,{\rm W m^{-2}\rm sr^{-1}}}
\def\twcm{\,{\rm 10^{20} cm^{-2}}}
\def\cm2{\,{\rm cm^{-2}}}
\def\kkms{\,{\rm K km s^{-1}}}
\def\sun{\odot}

\def\aua{{\rm A\&A} } 
\def\auas{{\rm A\&AS} } 
\def\apj{{\rm ApJ} } 
\def\aj{{\rm AJ} } 
\def\apjs{{\rm ApJS} } 
\def\apjl{{\rm ApJL} } 
\def\araa{{\rm ARAA} } 
\def\mnras{{\rm MNRAS} } 
\def\pasp{{\rm PASP} } 
\def\pasj{{\rm PASJ} }

\begin{document} 
 
\thesaurus{03(11.09.1 LMC; SMC;11.09.4;11.09.5;11.13.1;13.09.3;09.13.2)} 

\title{$\h2$ and its relation to CO in the LMC and other magellanic 
irregular galaxies} 
 
 
\author{F.P. Israel
        \inst{} 
  } 

 
\institute{Sterrewacht Leiden, P.O. Box 9513, 2300 RA Leiden, the Netherlands 
}  
 
\date{Received ????; accepted ????} 
 
\maketitle 

\begin{abstract} 

$\h2$ column densities towards CO clouds in the LMC and SMC are estimated
from their far-infrared surface brightness and HI column density. The newly 
derived $\h2$ column densities imply {\it N(H$_{2}$)}/{\it I(CO)} conversion 
factors (in units of 10$^{21}$ mol cm$^{-2}$ ($\kkms$)$^{-1}$) $X_{\rm LMC}$ 
= 1.3$\pm$0.2 and $X_{\rm SMC}$ = 12$\pm$2. LMC and SMC contain total 
(warm) $\h2$ masses of 1.0$\pm$0.3 $\times$ 10$^{8}$ $M_{\odot}$ and 
0.75$\pm$0.25 $\times$ 10$^{8}$ $M_{\odot}$ respectively. Local 
$\h2$/HI mass ratios similar to those in LMC and SMC are found in the 
magellanic irregulars NGC 55, 1569, 4214, 4449 and 6822 
and in the extragalactic HII region complexes NGC 604, 595 and 5461 in
M 33 and M 101 respectively. In these HII regions and in NGC 4449, we 
find {\it X} = 1--2; in NGC 55, 4214 and 6822 {\it X} = 3--6 
again in units of 10$^{21}$ mol cm$^{-2}$ ($\kkms$)$^{-1}$. The 
post-starburst galaxy NGC 1569 has a very high value similar to that of 
the SMC.

The CO--$\h2$ conversion factor {\it X} is found to depend on both the 
ambient radiation field intensity per nucleon $\sfir$/$\nh$ and 
metallicity [O]/[H]: 
log {\it X} $\propto$ 0.9$\pm$0.1 log$\frac{\sfir}{\nh}$ 
- 3.5$\pm$0.2 log$\frac{[O]}{[H]}$.
Neglecting dependency on radiation field, a reasonable approximation
is also provided by log {\it X} $\propto$ -2.7$\pm$0.3 log$\frac{[O]}{[H]}$. 
Milky Way values are consistent with these relations. 
This result is interpreted as the consequence of selective 
photodissociation of CO subjected to high radiation field energy densities 
and poor (self)shielding in low-metallicity environments, and especially the
preferential destruction of diffuse CO in `interclump' gas. 

Although locally $\h2$ may be the dominant ISM-component, the average 
global $\h2$/HI mass ratio is 0.2$\pm$0.04 and the average $\h2$ gas 
mass fraction is 0.12$\pm$0.02. Magellanic irregulars have warm molecular gas 
fractions very similar to those of our Galaxy, whereas other global 
properties (mass, luminosity, metallicity, CO luminosity) are very different.


\end{abstract} 

\section{Introduction} 

\subsection{$\h2$ content of galaxies}

The existence of molecular hydrogen ($\h2$) in interstellar space was 
suggested as early as 60 years ago by Eddington (1937) and Str\"omgren
(1939). Thirty years later, Gould $\&$ Salpeter (1963) and Hollenbach 
et al. (1971) predicted that it could be a large fraction of all
hydrogen. However, $\h2$ is difficult to observe directly, because it is 
a symmetrical molecule lacking a dipole moment. Nevertheless, it has 
been observed in absorption at UV wavelengths and in emission at infrared 
wavelengths. Because the emission arises mostly in warm or hot molecular 
gas, it has been virtually impossible to deduce total amounts of $\h2$ 
which is expected to be present mostly at low temperatures.

As $\h2$ is an abundant and important constituent of the interstellar 
medium in galaxies, there has been great interest in determining its 
presence and properties. Because it is commonly assumed that star 
formation requires interstellar clouds to pass through a cool, 
high-density phase in which most of the hydrogen is in molecular form, 
studies of star formation in external galaxies also seek to determine 
$\h2$ amounts in such galaxies.

Emission from the tracer CO molecule has been and still is widely used 
to determine the distribution and amount of molecular hydrogen in our 
own Milky Way and other galaxies. Although the usually optically thick
$^{12}$CO emission does not provide direct information 
on column densities, empirical relations between CO luminosities 
and virial masses of molecular clouds in the Galaxy suggest that 
circumstances nevertheless allow its use in an indirect manner. The 
underlying thought is that CO is distributed in a very clumpy manner, 
and that clumps are not self-shadowing. The strength of the signal received 
from many clumps in a single observing beam thus provides a measure for 
their total projected area weighted by brightness temperature. If the 
clumps are statistically similar from one line of sight to the other, we thus 
have a measure for the number of clumps per beam area, hence for the total 
amount of molecular material. In fact, the high optical depth of $^{12}$CO 
emission is then a boon, as it makes such determinations to first 
approximation independent of the actual [CO]/[$\h2$] abundance.

\subsection{Problems with CO-based methods}

The most commonly used methods to estimate molecular hydrogen contents of
extragalactic objects are either application of a `standard' CO to H$_{2}$
conversion factor {\it X} (defined as the ratio of molecular hydrogen column 
density {\it N(H$_{2}$)} to velocity-integrated CO intensity {\it I(CO)}) 
derived from Milky Way observations, or application of the virial theorem to 
observed CO clouds. The first method assumes similarity of extragalactic 
molecular clouds and Galactic clouds, or at least that the effects of 
different physical conditions cancel one another. In environments that are 
very different from those in the Galaxy, such as those found in galaxy central 
regions or in low-metallicity dwarf galaxies, this method must be considered
unreliable ({\it cf.} Elmegreen et al. 1980; Israel 1988; Maloney $\&$
Black 1988; Elmegreen 1989; Maloney 1990a). 
For instance, application of this method to the very low CO luminosities 
commonly observed for irregular dwarf galaxies would suggest negligible 
amounts of $\h2$ (Israel et al. 1995) and consequently unusually high star 
formation efficiencies (Israel 1997). In contrast, direct evidence for {\it X} 
factors varying by more than an order of magnitude, probably as a function of 
local conditions, has been presented for Galactic clouds by Magnani $\&$ 
Onello (1995). We thus agree with Roberts $\&$ Haynes (1994): `values of 
the molecular hydrogen content in late-type  systems derived in this 
manner are uncertain and possibly too low by up to an order of magnitude' 

The second method frequently used estimates total molecular cloud mass
from observed parameters such as CO extent {\it R(CO)} and velocity dispersion 
d{\it v(CO)}. Although this method, not assuming similarity between Galactic 
and extragalactic clouds, is preferable, it is likewise beset by problems, 
as it requires correct determination of the structure and dynamics of 
the observed clouds. For instance, the value of the virial constant used 
to convert observed parameters into mass may vary by a factor of four 
depending on the assumed condition of the system (see e.g. McLaren et 
al. 1988; McKee $\&$ Zweibel 1992), while it is not clear that the virial 
theorem is in fact relevant. If one considers the morphology 
of molecular complexes such as the ones in Orion (Bally et al. 1987), 
Taurus (Ungerechts $\&$ Thaddeus 1987) or indeed in the LMC (Israel 
$\&$ de Graauw 1991; Kutner et al. 1997) it is hard to 
imagine that these very elongated structures with little systematic 
velocity structure actually represent virialized complexes. Maloney 
(1990b) has shown that the correlation between CO luminosities and 
virial masses of Galactic molecular clouds follows directly from the 
size-linewidth relationship exhibited by molecular clouds and does 
not require virial equilibrium at all. Molecular hydrogen masses have 
also been determined applying {\it X}-factors scaled from $X_{\rm Gal}$ 
by {\it L(CO)} as a function of d{\it v} (e.g. LMC -- Cohen et al. 1988; SMC --
Rubio et al. 1991). 

Especially in the large linear beamsizes typical of extragalactic 
observations, actually unrelated clouds at somewhat different velocities 
may blend together, leading to unrealistical values of both cloud 
complex radius {\it R} and velocity dispersion d{\it v}. The derived (virial) 
masses may then either overestimate or underestimate the actual mass, 
depending on circumstances. For instance, consider an area mapped 
with a large beam blending together {\it N unrelated} clouds, each 
having a true mass $M_{\rm vir}$ = {\it a r} d$v_{\rm o}^{2}$. 
Here, $r_{\rm o}$ is the diameter of a single cloud and d$v_{\rm o}$ 
its velocity dispersion. The true total mass is thus {\it N a} $r_{\rm o}$ 
d$v_{\rm o}^{2}$. Cloud emission is measured over an area with radius 
{\it R} = $N^{1/2}$ {\it b} $r_{\rm o}$ in which {\it b} $r_{\rm o}$ is the 
projected separation between cloud centers. Unjustified application of the 
virial theorem on this observation suggests a total mass $M_{\rm vir}$ 
= {\it a} $N^{1/2}$ {\it b} $r_{\rm o}$ d$v_{\rm 1}^{2}$, where d$v_{\rm 1}$ 
now is the dispersion derived from the velocity width of the sum profile of 
all clouds within radius {\it R}. The ratio of the derived 
mass over the true mass is thus:

\begin{flushleft}
$M_{\rm derived}$/$M_{\rm true}$ = {\it b} $N^{-1/2}$ 
(d$v_{\rm 1}$/d$v_{\rm o}$)$^{2}$ \hfill (1)\\
\end{flushleft}

If {\it N} $<$ $b^{2}$ and d$v_{\rm 1}$ $>$ d$v_{\rm o}$, this will result in
a potentially large overestimate of the mass. However, if instead the 
unrelated clouds are at more or less identical radial velocities, 
d$v_{\rm 1}$ $\approx$ d$v_{\rm o}$, the true mass is underestimated 
if {\it N} $>$ $b^{2}$. Such a situation may occur in low-metallicity dwarf
galaxies with relatively small velocity gradients. It occurs if we have
a large number of clouds with small projected distances; a more physical 
equivalent is a very filamentary structure of the molecular material. 

A further problem in estimating H$_{2}$ masses from 
CO observations is the need to assume virtually identical 
distributions for both. If CO is significantly depleted, 
$\h2$ may well occur outside the area delineated by CO emission and its 
amount is underestimated by the CO measurements. This effect appears to lie 
at the base of the size dependence of {\it N(H$_{2}$)}/{\it I(CO)} ratio, 
noted by Rubio et al. (1993) and Verter $\&$ Hodge (1995). In 
low-metallicity galaxies suffering CO depletion, this results in a lack 
of CO emission in complexes observed on large angular scales. 
Observations on small angular scales selectively concentrate on the 
densest molecular components, that have resisted CO depletion most 
effectively, so that the {\it N(H$_{2}$)/I(CO)} ratio looks progressively 
more `normal' notwithstanding the lack of CO in {\it most} of the complex.

Thus, in order to estimate H$_{2}$ content of such galaxies, it is
desirable to use a method that does not require specific assumptions
on or knowledge of the detailed structure and dynamics of the molecular
clouds involved. Use of far-infrared data in principle provides such
a method (e.g. Thronson et al. 1987, 1988; Israel 1997).

\section{Method and data} 

\subsection{Estimating N(H$_{2}$) from $\sfir$ and N(HI)}

$\h2$ column densities are derived in the manner used on NGC 6822 by 
Israel 1997. At locations well away from star-forming regions and CO 
clouds, the ratio of neutral hydrogen column density to far-infrared 
surface brightness ({\it N(HI)/$\sfir$})$_{\rm o}$ is determined. In 
the absence of molecular gas, ({\it N(HI)/$\sfir$})$_{\rm o}$ equals 
{\it N$_{\rm H}$/$\sfir$}, which is a measure for the ambient gas-to-dust 
ratio. The observed $\sfir$ values at locations that contain $\h2$, as 
betrayed by CO emission, reduced to the reference dust temperature 
$\tdo$ and then multiplied by {\it (N(HI)/$\sfir$})$_{\rm o}$ thus
provide the total hydrogen column density $\nh$. The actual gas-to-dust 
ratio, which depends on poorly known dust particle properties, does not
need to be known as long as it does not change from source to reference
position. In the small irregular galaxies considered here, 
abundance gradients are negligible (cf. Vila-Costas $\&$ Edmunds 
1992), so that we may safely assume no change in gas-to-dust ratio as a 
function of position in the galaxy. When $\nh$ is known, {\it N(H$_{2}$)} is 
found by subtracting the local {\it N(HI)} value:

\begin{flushleft}
2 {\it N(H$_{2}$)} = [({\it N(HI)}/$\sfir$)$_{\rm o}$ {\it f(T)} $\sfir$] - {\it N(HI)} 
\hfill (2)\\
\end{flushleft}

In eqn. 2, {\it f(T)} is a function which corrects $\sfir$ for the 
emissivity difference due to the (generally small) difference of $\td$ 
from $\tdo$; for small temperature differences {\it f(T)} is close to
($\tdo$/$\td$)$^{6}$. Temperatures $\td$ are derived from the IRAS 
60$\mu$m/100$\mu$m flux ratio assuming a wavelength dependence for 
emission $I_{\lambda}$ $\propto$ $\lambda ^{-n} B _{\lambda}$. Here 
and in the following we will assume n = 2. The temperature correction 
assumes that the number distribution of dust particles emitting at varying 
temperatures does not differ significantly from one location to another. 
This is a reasonable assumption for values of {\it f(T)} not too far from 
unity, but may introduce significant errors for very large or very small 
values of {\it f(T)}. The CO to $\h2$ conversion factor {\it X} follows from 
the observed CO strength: {\it X} = {\it N(H$_{2}$)/I(CO)}. 

This method of estimating $\h2$ column densities depends on observed 
quantities {\it independent of the actual spatial or kinematical 
distribution of the molecular material}. It has this property in common 
with the methods used by Bloemen et al. (1986) and Bloemen et al. (1990) 
to estimate the same quantities in the Milky Way galaxy. 
It avoids the major weakness of the virial method discussed above,
as there is no need to determine the structure of the molecular cloud 
complexes, to separate unrelated clouds in the same line of sight, 
or even to resolve the molecular clouds. It is important to emphasize that
in this method, the absolute gas-to-dust ratio plays no role, nor does the
the actual dust mass. We thus avoid a major uncertainty associated with 
other infrared-derived $\h2$ estimates, where the infrared flux is used to
calculate a dust mass, which is then converted into a gas mass. Likewise,
our results are independent of CO measurements, and as we will show below, 
the observational uncertainties are no worse than those associated with 
the traditional methods and probably better.

\begin{table*}[htb]
\caption[]{LMC Data (unit area 0.043 kpc$^{2}$)}
\begin{flushleft} 
\begin{tabular}{lcccccccccl} 
\hline\noalign{\smallskip}
CO  & $\sfir$   & {\it f(T)} & {\it N(HI)} & {\it N(H$_{2}$)}  & $\frac{2 \nH2}{N(HI)}$ & $\frac{2 \nH2}{N_{\rm gas}}$ & {\it I(CO)$^{a}$} & {\it X} = $\frac{\nH2}{I(CO)}$ & $\frac{\sfir}{\nh}^{b}$ & Associated \\
Cloud & 10$^{-6}$ &    & \multicolumn{2}{c}{10$^{21}$} & 	         	   &	          		   &		 & 10$^{21}$        	    & 10$^{-28}$      		   & HII Region $^{c}$ \\ 
    & $\wmsr$   &      & \multicolumn{2}{c}{$\cm2$}  &             	   &	  			   & $\kkms$     & $\cm2 / \kkms$   	    & $\wmsr$cm$^{2}$ 		   & \\
\noalign{\smallskip}
\hline\noalign{\smallskip}
 3  &  6.0	& 0.38 &     0.7   &  2.6$\pm$0.7  & 7.4  & 0.65 & 0.75    & 3.4$\pm$1.1      & 10.3 	& N83B \\
 6  & 12	& 0.26 &     1.5   &  2.4$\pm$0.8  & 3.2  & 0.56 & 1.15    & 2.1$\pm$0.9      & 17.2 	& N11 \\
 6  &  1.2	& 1.00 &     1.2   &  0.8$\pm$0.3  & 1.3  & 0.42 & 1.35    & 0.6$\pm$0.2      &  4.4  	& N11-North \\
11  &  1.2	& 1.00 &     1.1   &  0.8$\pm$0.3  & 1.5  & 0.44 & 0.75    & 1.1$\pm$0.7      &  4.5     & (Bar)  \\
13  &  4.8	& 0.38 &     1.3   &  1.4$\pm$0.5  & 2.1  & 0.51 & 0.75    & 1.9$\pm$0.9      & 11.7	& N105A \\
15  &  2.4	& 0.57 &     1.0   &  1.0$\pm$0.4  & 2.0  & 0.49 & 1.15    & 1.0$\pm$0.5      &  7.8 	& N113 \\
--  &  1.8	& 0.57 &     1.2   &  0.6$\pm$0.3  & 1.0  & 0.37 & 0.40    & 1.4$\pm$1.1      &  7.8	& N30 \\
18  &  2.1	& 1.00 &     0.9   &  1.9$\pm$0.6  & 4.2  & 0.60 & 1.15    & 1.6$\pm$0.6      &  4.5	& (Bar) \\	
19  & 12	& 0.31 &     2.0   &  3.2$\pm$1.0  & 3.2  & 0.56 & 2.50    & 1.3$\pm$0.4      & 14.3	& N44 \\	 
20  &  1.8	& 0.51 &     1.5   &  0.3$\pm$0.3  & 0.4  & 0.21 & 1.15    & 0.3$\pm$0.3      &  8.7	& N18/N144 \\
22  &  1.2	& 0.57 &     0.7   &  0.4$\pm$0.2  & 1.1  & 0.40 & 0.75    & 0.6$\pm$0.3      &  7.8	& N132 \\
23  &  3.6	& 0.38 &     1.5   &  0.8$\pm$0.4  & 1.1  & 0.38 & 1.50    & 0.5$\pm$0.3      & 11.7	& N48 \\
26  &  4.2	& 0.38 &     1.0   &  1.3$\pm$0.4  & 2.6  & 0.53 & 0.40    & 3.3$\pm$1.8      & 11.7	& N206 \\
27  &  2.4	& 0.51 &     1.6   &  0.6$\pm$0.3  & 0.8  & 0.32 & 1.15    & 0.5$\pm$0.3      &  8.7	& N148/N150 \\	
29  &  4.5	& 0.38 &     1.1   &  1.4$\pm$0.5  & 2.6  & 0.53 & 0.40    & 3.6$\pm$1.9      & 11.7	& N57 \\
31  &  1.8	& 0.57 &     1.0   &  0.7$\pm$0.3  & 1.4  & 0.43 & 0.40    & 1.6$\pm$1.2      &  7.8	& N64 \\
32  & 120	& 0.08 &     3.2   &  9.7$\pm$2.7  & 6.1  & 0.64 & 1.15    & 8.4$\pm$2.9      & 52.9	& 30 Dor \\
32$^{d}$ & 9.0	& 0.26 &     2.5   &  1.4$\pm$0.6  & 1.1  & 0.39 & 1.50    & 0.9$\pm$0.4      & 17.1	& N157B \\
33$^{d}$ & 12   & 0.26 &     3.2   &  1.9$\pm$0.8  & 1.2  & 0.40 & 4.00    & 0.5$\pm$0.3      & 17.1	& N159 \\
34  &  1.2	& 1.20 &     3.0   &  0.1$\pm$0.4  & 0.1  & 0.05 & 0.55    & 0.2$\pm$0.6      &  1.9	& N214 \\
35  &  3.6	& 1.00 &     3.5   &  2.3$\pm$1.0  & 1.3  & 0.44 & 4.50    & 0.5$\pm$0.2      &  5.1	& N176 \\
36  &  1.8	& 1.45 &     3.0   &  1.4$\pm$0.7  & 0.9  & 0.36 & 2.30    & 0.6$\pm$0.3      &  3.1 	& N216 \\
37  &  4.8	& 0.57 &     3.0   &  1.6$\pm$0.7  & 1.1  & 0.38 & 2.60    & 0.6$\pm$0.3      &  7.8  	& N167 \\
\noalign{\smallskip}	
\hline
\noalign{\smallskip}	
Mean$^{e}$ & 4.3 & 0.61 &    1.7   &  1.3$\pm$0.2  & 1.9  & 0.42 & 1.40	  & 1.3$\pm$0.2      &  9.2$\pm$1.0 & --- \\
\noalign{\smallskip}	
\hline
\end{tabular}
\end{flushleft} 
Notes: a. Uncertainty in {\it I(CO)} is 0.24 $\kkms$ (Cohen et al. 1988); 
b. Uncertainty in $\frac{\sfir}{\nh}$ is 25$\%$; c. Henize 1956; 
d. Data unreliable because of location on strong emission gradients. 
e. Mean values excluding 30 Doradus.
\end{table*} 

The column densities {\it N(H$_{2}$)}, and consequently {\it X}, determined 
in this paper are properly lower limits (Israel 1997). {\it (i)}. If some 
$\h2$ were to be present at the null positions where we assumed none, the 
total hydrogen column density corresponding to unit infrared luminosity is 
underestimated, implying higher actual {\it N(H$_{2}$)} values than derived. 
{\it (ii)}. If, unexpectedly, the {\it hotter} infrared sources were to
be relatively rich in {\it cooler} dust, the observed infrared emission
does not sample the total amount of gas, hence {\it N(H$_{2}$)}, will be higher 
than estimated. 
{\it (iii)}. If, in regions of bright infrared emission, higher 
radiation densities would cause increased dust depletion, these regions
will be characterized by a higher gas-to-dust ratio than assumed, again 
leading to higher than derived actual {\it N(H$_{2}$)} values. This is expected 
only to be important for HII regions filling a significant fraction of 
the beam. 

Errors in the assumptions would thus cause {\it N(H$_{2}$)} and {\it X} to be 
higher rather than lower. Although these errors are hard to quantify, we 
consider it unlikely that their effect will exceed a factor of two. The 
calculated total hydrogen column densities $\nh$ carry with them the combined 
uncertainty in the determinations of ({\it N(HI)}/$\sfir$)$_{\rm o}$, 
{\it f(T)} and $\sfir$. Because these quantities are compared in a relative 
rather than an absolute sense, the uncertainty {\it $\Delta \nh$} is of 
the order of 20$\%$ - 30$\%$ for the cases discussed below. The uncertainty 
in the calculated values of {\it N(H$_{2}$)} is larger. Since the {\it N(HI)} 
determinations are considered to be rather accurate, it depends on the 
molecular to atomic hydrogen ratio: 
$\Delta N(H_{2})$ = $\Delta \nh$(1 + 0.5 {\it N(HI)}/{\it N(H$_{2}$)}) \\

Thus, for $\h2$ column densities equal to or higher than those observed
in HI, the relative $\h2$ uncertainty is typically less than 50$\%$.
For HI column densities substantially higher than the derived $\h2$ 
column density, the relative uncertainty may become considerable. 
However, this situation almost exclusively occurs at low absolute 
N($\h2$) values where a relatively large uncertainty still corresponds 
to an acceptable uncertainty in the absolute value. The uncertainty in 
the derived value of {\it X}, in turn, includes both the uncertainty in 
{\it N(H$_{2}$)} and in {\it I(CO)}. Since the latter is usually much smaller than 
the former, the uncertainty in {\it X} is actually dominated by that in 
{\it N(H$_{2}$)}. The combined effect of uncertainties in the observational 
values and in the assumptions implies a rough overall uncertainty of about a
factor of two for individual determinations. 

\subsection{Data and results}

All data were taken from the literature or existing databases. The CO, 
HI and far-infrared data included in the comparison are selected to have 
similar resolutions. This resolution is determined by the lowermost 
resolution to which the other data are degraded, if necessary.

\subsubsection{LMC}

The far-infrared data are from Schwering 1988, who conveniently 
produced maps of infrared luminosity over HI mass at 15$\arcmin$
resolution (corresponding to 235 pc) and dust temperature at 8$\arcmin$ 
resolution (Fig. 1). The HI data (resolution 15$\arcmin$) are from Rohlfs 
et al. 1984. The average of six positions in the main body of the LMC, 
well away from CO clouds and bright HII regions is 
({\it N(HI)}/$\sfir$)$_{\rm o}$ = 2.25$\times$10$^{27}$ cm$^{-2}$/$\wmsr$ 
(corresponding to {\it L}/{\it M} = 1.7 $L\sun$/$M\sun$) at a reference 
temperature $\tdo$ = 25.5 K. From the internal variation, we estimate its 
uncertainty to be about 10$\%$. The uncertainty in {\it f(T)} is about 
20$\%$ and that in $\sfir$ about 10$\%$. 

\begin{figure}
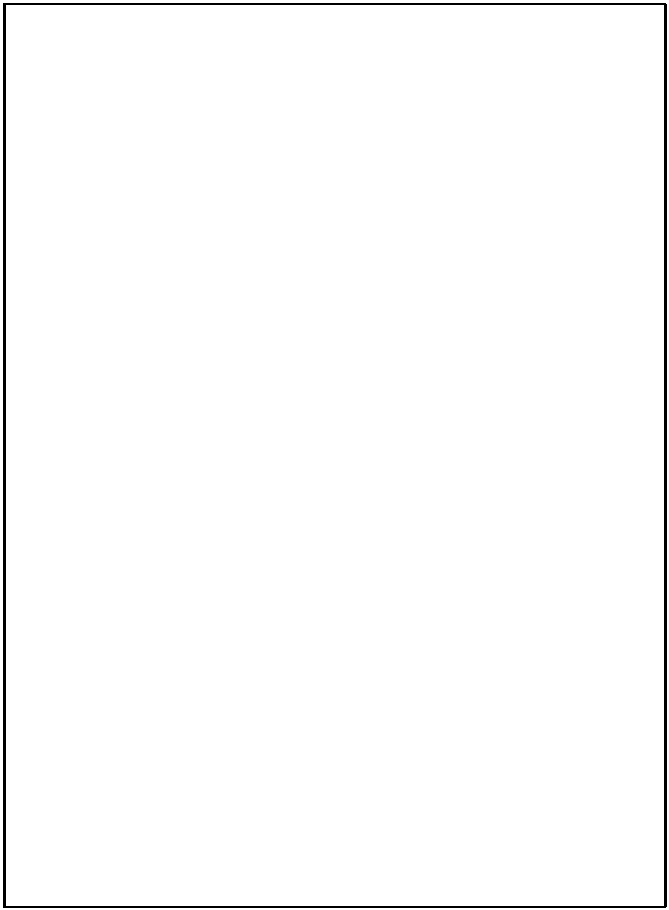
 
\picplace{12.0cm} 
\caption[]{LMC. Top: ratio of far-infrared luminosity to neutral
hydrogen column density at 15$\arcmin$ resolution. Contours
are at 5.3, 13, 26, 52, 105 $\times$ 10$^{-28}$ W m$^{-2}$ sr$^{-1}$
cm$^{2}$. Bottom: dust temperature $\td$ at 8$\arcmin$ resolution. 
Contours are at 23, 28, 32 and 36 K.
}            
\end{figure} 

In Table 1 we have listed data for several of the CO cloud complexes 
detected by Cohen et al. (1988) convolved to a resolution of 15$\arcmin$
(e.g. Meinert, 1992). Except for cloud 31, all CO clouds
considered have diameters larger than 15$\arcmin$). Weaker CO sources 
are included only if identification with an HII region complex 
support their validity. In Table 1, the first column identifies 
the CO cloud by its number in Table 1 of Cohen et al. (1988). Column 
2 lists the far-infrared surface brightness at the peak CO position, 
column 3 the value of {\it f(T)} based on the dust temperature 
derived from the $F_{60}$/$F_{100}$ flux ratio and column 4 the HI 
column density. Column 5 gives the molecular hydrogen column densities 
calculated according to eqn. 2, while columns 6 and 7 give the resulting 
ratios of molecular hydrogen to atomic hydrogen and total gas (including 
helium) respectively. Column 8 gives the integrated CO intensity and column 
9 the resulting value of {\it X}. In column 10 we give the ratio of the 
observed infrared surface brightness ({\it not} reduced in temperature) 
over total hydrogen column density $\nh$ = 2{\it N(H$_{2}$)} + {\it N(HI)}. 
This ratio is a measure of the ambient radiation field strength per H nucleon. 
Finally, column 11 lists HII region(s) associated with the molecular cloud. 
In most cases the HII region extent is much less than the 15$\arcmin$ scale 
relevant to the data used.

Some further comments are in order. Clouds 34, 35 and 36 are located 
south of the bright HII regions associated with the Doradus complex.  
Major CO emission occurs with little or no optical counterpart. 
There is relatively strong HI emission, but the far-infrared surface 
brightness decreases smoothly. The results for N157B and N159 (clouds 
32 and 33) are uncertain, as both are at steep far-infrared gradients. 
N159 is also on a steep CO emission gradient in the opposite direction. 
Consequently, the resulting value of $N(H_{2}$ depends critically on 
the precise (within a fraction of the resolution) position used.
The {\it mean} value of the CO to $\h2$ conversion ratio (excluding 
30 Doradus) is {\it X} =  13($\pm$2)$\times \twcm$. Its uncertainty 
is determined by that in ({\it N(HI)}/$\sfir$)$_{\rm o}$, which does 
not decrease with increasing sample size, whereas all other errors do. 

\begin{figure}
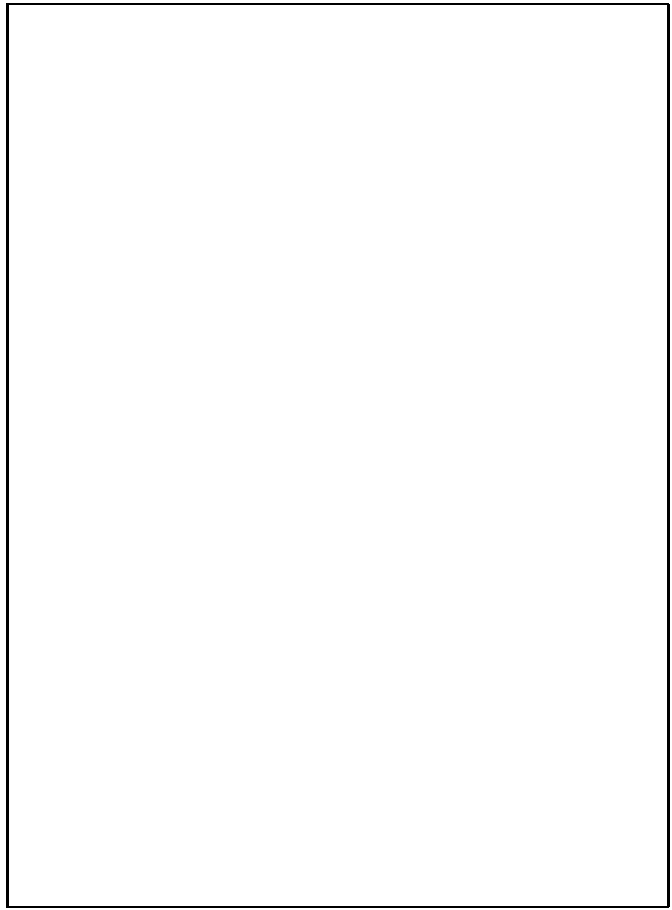
 
\picplace{12.0cm} 
\caption[]{SMC. Top: ratio of far-infrared luminosity to neutral
hydrogen column density at 15$\arcmin$ resolution. Contours
are at 2.6, 5.2, 10.5, 15.7, 21, 26 $\times$ 10$^{-29}$ W m$^{-2}$ 
sr$^{-1}$ cm$^{2}$. Bottom: dust temperature $\td$ at 8$\arcmin$
resolution. Contours are at 28, 32 and 36 K.
}            
\end{figure} 

\subsubsection{SMC}

\begin{table*}
\caption[]{SMC Data (unit area 0.061 kpc$^{2}$)} 
\begin{flushleft} 
\begin{tabular}{lcccccccccl} 
\hline\noalign{\smallskip}
CO  & $\sfir$   & {\it f(T)} & {\it N(HI)} & {\it N(H$_{2}$)}  & $\frac{2 \nH2}{N(HI)}$ & $\frac{2 \nH2}{N_{\rm gas}}$ & {\it I(CO)$^{a}$} & {\it X} = $\frac{\nH2}{I(CO)}$ & $\frac{\sfir}{\nh}^{b}$ & Associated \\
Cloud & 10$^{-6}$ &    & \multicolumn{2}{c}{10$^{21}$}    &         	   &	          		   &		 & 10$^{21}$        	    & 10$^{-28}$      		   & HII Region $^{c}$ \\ 
    & $\wmsr$   &      & \multicolumn{2}{c}{$\cm2$}       &             	   &	  			   & $\kkms$     & $\cm2 / \kkms$   	    & $\wmsr$cm$^{2}$ 		   & \\
\noalign{\smallskip}
\hline
\noalign{\smallskip}
SW1 & 1.5       & 0.74  &    7.8	&  6.3$\pm$2.8  & 1.6 & 0.46 & 0.42    & 15$\pm$7 	     & 0.7 	& N16/N17 \\
SW2 & 1.3	& 0.67  &    8.8        &  3.6$\pm$2.1  & 0.8 & 0.33 & 0.36    & 10$\pm$6         & 0.8 	& N25/N26 \\
 -  & 1.5	& 0.55  &    7.0        &  4.2$\pm$2.1  & 1.2 & 0.40 & 0.20    & 20$\pm$12        & 1.0  	& N32/N37 \\
NE1 & 0.3	& 0.90  &    3.0        &  1.3$\pm$0.8  & 0.9 & 0.34 & 0.20    &  7$\pm$4         & 0.6      & N72 \\
NE2 & 1.3	& 0.55  &    4.3	&  4.4$\pm$1.8  & 2.1 & 0.50 & 0.33    & 13$\pm$6         & 1.0	& N66 \\
NE3 & 0.9	& 0.67  &    5.9	&  2.5$\pm$1.5  & 0.9 & 0.34  & 0.30   &  8$\pm$5         & 0.8 	& N76 \\
\noalign{\smallskip}	
\hline
\noalign{\smallskip}	
Mean & 1.1      & 0.68 &     6.1	&  3.7$\pm$0.8  & 1.3 & 0.40 & 0.30   & 12$\pm$2         & 0.8$\pm$0.1 & --- \\
\noalign{\smallskip}
\hline
\end{tabular}
\end{flushleft} 
Notes: a. Uncertainty in {\it I(CO)} is 0.06 $\kkms$ (Rubio et al. 1991); 
b. Uncertainty in $\frac{\sfir}{\nh}$ is 25$\%$; c. Henize 1956.
\end{table*} 

\begin{table*}
\caption[]{Other Galaxies} 
\begin{flushleft} 
\begin{tabular}{lcccccccccr} 
\hline\noalign{\smallskip}

Galaxy       & $\sfir$   & {\it f(T)} & {\it N(HI)}   & {\it N(H$_{2}$)} & $\frac{2 \nH2}{N(HI)}$ & $\frac{2 \nH2}{N_{\rm gas}}$ & {\it I(CO)$^{a}$} & {\it X} = $\frac{\nH2}{I(CO)}$ & $\frac{\sfir}{\nh}$ & Unit \\
             & 10$^{-6}$ &      & \multicolumn{2}{c}{10$^{21}$} &      &      &         & 10$^{21}$       & 10$^{-28}$       & Area \\
             & $\wmsr$   &      & \multicolumn{2}{c}{$\cm2$}  &          &      & $\kkms$ & $\cm2 / \kkms$  & $\wmsr$cm$^{2}$  & kpc$^{2}$ \\
\noalign{\smallskip}
\hline
\noalign{\smallskip}
NGC 55       & 20	& 1.0  &   8.0  & 6.0$\pm$3.0 	    &  1.5     & 0.44 & 2.0    & 3.0$\pm$1.7      &  10$\pm$2    & 0.6 \\
NGC 1569     & 12	& 1.5  &   3.5  & 6.0$\pm$3.0       &  3.4     & 0.57 & 0.37   & 16$\pm$8         &   8$\pm$2    & 0.7 \\
NGC 4214     &  7	& 0.9  &   1.6  & 2.3$\pm$0.9       &  2.9     & 0.55 & 0.7    & 3.3$\pm$1.5      &  12$\pm$2    & 4.2 \\
NGC 4449     &  1.6     & 0.45 &   1.85 & 0.7$\pm$0.3	    &  0.6     & 0.28 & 0.95   & 8.0$\pm$3.0	  & 2.3$\pm$0.6  & 2.1 \\
\noalign{\smallskip}
\hline
\noalign{\smallskip}
NGC 6822-HII &  0.85	& 0.7  &   1.8  & 2.8$\pm$0.9       &  3.1     & 0.56 & 0.50   & 5.5$\pm$1.1      &  1.1$\pm$0.2 & 0.1 \\
NGC 6822-IR  &  0.75	& 0.8  &   1.5	& 3.0$\pm$0.5       &  4.0     & 0.59 & 0.85   & 4.7$\pm$1.3      &  1.0$\pm$0.2 & 0.1 \\
\noalign{\smallskip}
\hline
\noalign{\smallskip}
NGC 604      &  2.0	& 0.6  &   3.0	& 2.6$\pm$1.0       &  1.7     & 0.47 & 1.2    & 2.2$\pm$0.9      &  2.4$\pm$0.5 & 0.5 \\
NGC 595      & 1.1	& 1.0  &   1.9	& 1.0$\pm$0.4       &  1.1     & 0.38 & 0.9    & 1.2$\pm$0.9      &  2.7$\pm$0.6 & 0.5 \\
NGC 5461     &  6	& 1.0  &   1.6	& 2.7$\pm$0.9       &  3.4     & 0.57 & 2.5    & 1.1$\pm$0.4      &  8$\pm$2     & 3.4 \\ 
\noalign{\smallskip}
\hline
\end{tabular}
\end{flushleft} 
Note: a. For CO details, see text.
\end{table*} 

The far-infrared data are from Schwering's (1988) maps 
of dust temperature and of far-infrared luminosity 
over HI mass (Fig. 2). The HI data at the same resolution are
from McGee $\&$ Newton (1981). For the SMC we find an average 
({\it N(HI)}/$\sfir$)$_{\rm o}$ = 1.65($\pm$0.25)$\times$10$^{28}$ 
cm$^{-2}$/$\wmsr$ (corresponding to {\it L}/{\it M} = 0.23 $L\sun$/$M\sun$) 
for various positions in and near the bar, at a reference temperature 
$\tdo$ = 28 K. In Table 2 we list the SMC data for several of 
the CO cloud complexes detected by Rubio et al. (1991) in the same
format as Table 1.

The {\it mean} value of the CO to $\h2$ conversion ratio is {\it X} 
= 120($\pm$30)$\times \twcm$. The uncertainty in the null determination
again dominates, but less decisively because of the relatively small 
sample size in Table 2. 

\subsubsection{NGC 55, NGC 1569, NGC 4214 and NGC 4449}

Four other irregular galaxies have far-infrared, HI and CO data at similar 
resolutions (Table 3). For these galaxies, we used far-infrared data at 
a resolution of 1.4$\arcmin$ obtained with IRAS CPC instrument at 
50$\mu$m and 100$\mu$m (F. Sloff, unpublished; Van Driel et al. 1993). 
For consistency, we interpolated the CPC 50$\mu$m fluxes to 60$\mu$m; as
the absolute calibration of the CPC is unreliable (Van Driel et al. 1993),
we scaled all CPC fluxes by the IRAS survey fluxes. In the case of NGC 55 
we verified the outcome by comparison with the IRAS survey image-sharpening 
(PME) result published by Bontekoe et al. (1994). 

{\it NGC 55} was sampled at the CO cloud detected by Dettmar $\&$ Heithausen 
(1989) and at two reference positions 1.5$\arcmin$ on either side of 
this peak. Using HI data from Hummel et al. (1986), we found for the 
reference value ({\it N(HI)}/$\sfir$)$_{\rm o}$ = 
1.4$\pm$0.2$\times$10$^{28}$ cm$^{-2}$/$\wmsr$. Dettmar $\&$ Heithausen (1989) 
give a CO surface brightness of 3 $\kkms$ and a source size of 100$\arcsec 
\times$40$\arcsec$, so that {\it I(CO)} = 2 $\kkms$ in a 
1$\arcmin$ beam.  

{\it NGC 1569} was observed in CO by Greve et al. (1996) who detected
a cloud with a peak I(CO) = 2.9 $\kkms$ and a size of 22$\arcsec$,
corresponding to a CO intensity of 0.37 $\kkms$ in a 1$\arcmin$ beam. 
HI data at 1$\arcmin$ resolution are from J. Stil (private communication; 
see also Israel $\&$ Van Driel 1990). At the reference position 1.2$\arcmin$ 
to the southeast, where Greve et al. (1996) did not find CO emission, we 
determined ({\it N(HI)}/$\sfir$)$_{\rm o}$ = 7.7$\times$10$^{28}$ 
cm$^{-2}$/$\wmsr$. Infrared emission gradients render this result uncertain
by 25$\%$. As weak CO emission might be present outside the limited 
area mapped, the uncertainty in {\it X} may be as high as 65$\%$. If we take
the Young et al. (1984) results ({\it I(CO)} = 1.1$\pm$0.3 $\kkms$ in a 
50$\arcsec$ beam), we would find an {\it X}-ratio only half the value 
in Table 3, which we take as indicative of the uncertainty in {\it X}. 

{\it NGC 4214} was observed in CO by Becker et al. (1995) who found a 
cloud complex of dimensions 36$\arcsec \times$22$\arcsec$ with a
peak {\it I(CO)} = 2.8 $\kkms$ in a 21$\arcsec$ beam. After correction
for beamsize, this is consistent with earlier and less accurate 
determinations by Tacconi $\&$ Young (1985) and Thronson et al. (1988). 
Weak CO was also detected by Ohta et al. (1993) in a 15$\arcsec$ beam 
towards positions 45$\arcsec$ southeast and 35$\arcsec$ northwest of 
the reference position given by Becker et al. (1995). On the basis of 
all available data, we take {\it I(CO)} = 0.7($\pm$0.15) $\kkms$ in 
a 1$\arcmin$ beam. HI data at 1$\arcmin$ resolution are from Allsop (1979). 
At two reference positions, we determined ({\it N(HI)}/$\sfir$)$_{\rm o}$ = 
1.2$\pm$0.2$\times$10$^{27}$ cm$^{-2}$/$\wmsr$. 

{\it NGC 4449} was observed in CO by Hunter $\&$ Thronson 1996 (65$\arcsec$
beam) and by Sasaki et al. 1990 (15$\arcsec$ beam). Taking into account 
beam sizes and efficiencies, the data agree well. HI data at
1$\arcmin$ resolution are from the WHISP database (J. Kamphuis, private 
communication). We obtained reasonably accurate infrared surface 
brightnesses for Hunter $\&$ Thronson's regions 1 through 4 only and 
determined ({\it N(HI)}/$\sfir$)$_{\rm o}$ = 
4.35$\pm$1.0$\times$10$^{27}$ cm$^{-2}$/$\wmsr$. Table 3 gives the mean 
of the individual results for the four positions, weighted by {\it I(CO)}.

\subsubsection{Extragalactic HII regions}

In Table 3, we have also included the results obtained for NGC 6822 
(Israel 1997). Two sets of entries are given: NGC 6822-HII represents 
the mean values towards the HII region complexes Hubble I/III, V and X, 
and NGC 6822-IR those towards the infrared sources 4 and 6 not associated 
with major HII regions.

We also included data for the bright HII regions NGC 604 and NGC 595 in
M 33. The infrared data were taken from Rice et al. (1990), HI data at
the same resolution from Deul (1988) and CO data from Blitz (1985). 
Note that the results for NGC 595 are rather
uncertain, as a reliable flux at 100$\mu$m is hard to determine; we
assumed essentially similar infrared flux distribtions for both NGC 604
and NGC 595. Finally, we also included NGC 5461, the brightest HII region 
in the galaxy M 101. Infrared data were taken from Bontekoe et al. (1994),
HI data from van der Hulst $\&$ Sancisi (1988) and CO data from Blitz et 
al. (1981) and Kenney et al. (1991). In this case, we applied 
a correction factor of 1.3 to allow for the presence of significant amounts 
of HII; this effect is negligible for the M 33 objects. Because the parent 
galaxies M 33 and M 101 have radial abundance gradients, we selected null 
positions adjacent to the HII regions used.

\section{Analysis $\&$ Discussion} 

\subsection{The CO to H$_{2}$ conversion factor}

In the sample galaxies, we find CO to $\h2$ conversion factor {\it X} values
much higher than the range 0.2-0.4 $\times$ 10$^{21}$ $\cm2$ 
($\kkms$)$^{-1}$ commonly adopted for Milky Way objects. In the LMC, we 
find a mean value {\it X} = 1.3 $\times$ 10$^{21}$ $\cm2$ ($\kkms$)$^{-1}$, 
or 3 - 7 times higher than in the Milky Way. We may compare this result
to that obtained by Cohen et al. (1988). Comparing CO luminosities
{\it L(CO)} to velocity width $\Delta v$, they conclude that on 
average $X_{\rm LMC}$ = 6$X_{\rm G}$. They adopt $X_{\rm G}$ 
= 0.28 $\times$  10$^{21}$ $\cm2$ ($\kkms$)$^{-1}$ resulting in a value for 
$X_{\rm LMC}$ 30$\%$ higher than ours. However, if $X_{\rm G}$ 
= 0.20 $\times$ 10$^{21}$ $\cm2$ ($\kkms$)$^{-1}$ (Bloemen 1989), their 
result is identical to ours. 
Satisfactory as this may seem, the situation is more complex. 

First, Cohen et al. (1988) estimated their value of {\it X} from the mean ratio 
$L(CO)_{\rm G}$/$L(CO)_{\rm LMC}$, but their Fig. 2 shows this ratio to 
be a function of $\Delta v$. At the smallest velocity widths, their 
implied $X_{\rm LMC}$ is only 3 $X_{\rm G}$, but at the largest widths 
it is 10$X_{\rm G}$. The figure exhibits a large scatter around the 
mean, covering an equivalent range in $X_{\rm LMC}$ of 2 - 20 $X_{\rm G}$. 
Part of this scatter is undoubtedly due to the low signal-to-noise 
ratio of the CO measurements, but part of it is real. For instance, 
Garay et al. (1993) studied seven CO clouds in the 30 
Doradus halo and found those to have $L(CO)_{\rm G}$/$L(CO)_{\rm Dor}$ 
ratios implying a much higher value $X_{\rm Dor}$ = 20 $X_{\rm G}$ than 
the mean found by Cohen et al. (1988). We note that the values tabulated 
in our Table 2 also define a large range in {\it X}, from close to the 
Galactic value to more than an order of magnitude higher. For 30 Doradus 
itself we derive an even higher {\it X} value. We will discuss this variation 
in {\it X} in sect. 3.2.

\begin{table}
\caption[]{Comparison of X Values} 
\begin{flushleft} 
\begin{tabular}{lccl} 
\hline\noalign{\smallskip}
Galaxy       & \multicolumn{2}{c}{\it X}	    & References \\
             & \multicolumn{2}{c}{10$^{21} \cm2 / \kkms$}        & \\ 
	     & This Paper       & Literature	    & \\
\noalign{\smallskip}
\hline
\noalign{\smallskip}
LMC	     & 1.3		& 1.7; 3.9 	& 1; 2 \\
SMC	     & 12 		& 6 		& 3 \\
NGC 55       &  3		& 6 		& 4 \\
NGC 1569     & 16		& 5 		& 5 \\
NGC 4214     &  3		& 2; 1 		& 6; 7 \\
NGC 4449     & 0.8              & 1		& 8 \\
NGC 6822     &  5		& 0.6--3	& 9 \\
NGC 604      & 2.2		& 3; 0.35; 3.5 	& 10; 11, 12; 13 \\
NGC 595      & 1.2		& 1.4; 0.6 	& 10; 12 \\
NGC 5461     & 1.1		& 2 		& 14 \\ 
\noalign{\smallskip}
\hline
\end{tabular}
\end{flushleft} 
References for other {\it X} values: 1. Cohen et al. (1988); 2. Garay et al. 
(1993); 3. Rubio et al. (1991); 4. Dettmar $\&$ Heithausen (1989); 
5. Greve et al. (1996); 6. Thronson et al. (1988); 7. Becker et al. 
(1995); 8. Estimated from Ohta et al. 1993, after correction for main-beam
efficiency; 9. Wilson (1994); 10. Estimated from Blitz (1985); 11. Viallefond
et al. (1992); 12. Wilson $\&$ Scoville (1992); 13. Israel et al. (1990); 
14. Estimated from Blitz et al. (1981).
\end{table} 

Second, we differ in individual cases, even though the mean values 
agree. For instance, Cohen et al. (1988) obtain very high $\h2$ and 
virial masses (and their Fig. 2 implies a high value for {\it X}) in 
cloud 35. In this cloud, we find a {\it low X} value. The result by 
Cohen et al. (1988) follows from the high $\Delta v$ = 28 km s$^{-1}$ 
they found for this cloud. They list similarly large velocity 
widths for e.g. clouds 13, 19 and 23. Yet, the higher resolution SEST 
survey yields velocity widths typically a factor of two less (Israel 
et al. 1993; Kutner et al., 1997). At least in the case of cloud 35, 
the anomalously high velocity width appears to be the result of CO clouds 
at two distinct velocities blended together in the large beam used by Cohen 
et al. (1988). Reduction of the large velocity widths to the more modest 
SEST values, yields {\it X} values in much better agreement with those in 
Table 2. Similar comments apply to the SMC results by Rubio et al.
(1991), except that here we find a {\it larger} value of {\it X}, although
both results have significant uncertainties associated with them. 
It is nevertheless clear that $X_{\rm SMC}$ is much higher 
than either $X_{\rm LMC}$ or $X_{\rm G}$. 

For the objects listed in Table 3, {\it X} estimates can be obtained from
the literature (Table 4). These are mostly rough estimates based on 
comparison of CO luminosities and virial masses, and are very uncertain. 
Nevertheless, we see reasonably good agreement in Table 4 for 
NGC 55, NGC 4214, NGC 4449 and for NGC 5461. There is also good 
agreement for NGC 595 and NGC 604 if we disregard the estimates 
derived from the high resolution observations which apply to individual 
cloud components rather than whole complexes. It has been noted before 
(Rubio et al. 1993; Verter $\&$ Hodge, 1995) that such observations 
always yield {\it X} values much lower than the global values derived from 
observations covering the whole complex (see als sect. 1.2). 
Our results for NGC 6822, and especially for NGC 1569, are higher 
than the other published estimates.

\subsection{Dependence of X on environment}

\subsubsection{Dependence on radiation field}

Compared to the Milky Way, the galaxies studied here have lower
metallicities, and are found to have higher {\it X} ratios. This is
not a new result: several authors have suggested a more or less
inverse linear dependence between {\it X} and metallicity as measured by
the oxygen abundance ({\it cf.} Dettmar $\&$ Heithausen, 1989; Rubio 
et al. 1991; Verter $\&$ Hodge 1995; Arimoto 1996, Sakamoto 1996). Arnault 
et al. (1988) found the ratio of CO to HI emission in a sample of 19 late-type 
galaxies to vary as roughly as [O]/[H]$^{2.2}$. Sage et al (1992) failed 
to reproduce such a relation, but our sample suggests the CO to HI ratio
to be proportional to [O]/[H]$^{2.6}$, within the errors identical
to the result obtained by Arnault et al. (1988). The latter conclude 
that CO must be deficient relative to $\h2$ as a function of 
metallicity, but could not determine a functional dependence for
the CO/$\h2$ ratio, hence {\it X}, on metallicity. Their result
nevertheless suggests a dependence with a coefficient larger than unity,
as there is no reason to assume vastly different $\h2$/HI ratios in
low-metallicity galaxies. The present data provide an excellent basis to
pursue this question, as they have been obtained in a consistent manner.
Much of the previously published discussions were based on a compilation
of data (notably {\it X}-values) from different sources, and obtained in
different manners.

\begin{figure}
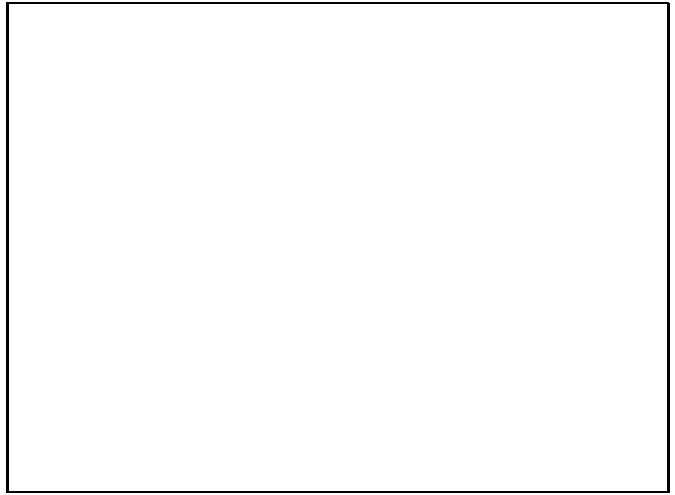
 
\picplace{6.5cm} 
\caption[]{Dependence of {\it X} on radiation field as represented by the ratio
$\sfir$/$\nh$. Left: LMC; right SMC. Linear regression lines are plotted.
}            
\end{figure} 

An underabundance of CO with respect to hydrogen is expected to result
from low carbon and oxygen abundances. It will be enhanced by 
photodissociation of CO since low metal abundances imply both lower
selfshielding and lower dust shielding of CO against the ambient
radiation field. The LMC and SMC samples allow us to first investigate the
effect of the radiation field. In Fig. 3 we have plotted the ratio 
{\it N(H$_{2}$)/I(CO)} = {\it X} as a function of the energy available per H 
nucleon, represented by the quantity $\sfir$/$\nh$. Straight lines indicate 
the linear regression lines. In case of the LMC, we did not include 
30 Doradus in determining the regression line, because its very
high values would dominate the result. Nevertheless, extrapolation of
the regression line to the $\sfir$/$\nh$ value of 30 Doradus predicts it
to have {\it X} = 7 $\times \twcm \cm2$ ($\kkms$)$^{-1}$, or 85$\%$ of the
value derived directly in Table 1. The SMC sample, although much smaller,
shows a similar behaviour. Further analysis suggests that the dependence
of X on radiation field is indeed very close to linear: {\it X} $\propto$ 
($\sfir$/$\nh$)$^{0.9\pm 0.1}$.

The increase of {\it X}, i.e. the decrease of CO relative to {\it N(H$_{2}$)},
as more energy per nucleon is available is the result of two processes, 
as is illustrated by the SEST results obtained for the LMC. High-resolution
maps (linear beamsize corresponding to 10 pc) were obtained of clouds 
35/36 (south of 30 Doradus -- Kutner et al. 1997), cloud
6 (N11 -- see Israel $\&$ de Graauw 1991) and cloud 32 (30 Doradus --
Johansson et al., in preparation). The map of cloud 35/36 shows numerous 
clumps embedded in extended interclump gas; the average peak-to-diffuse 
CO contrast ratio is about 3. Bright clumps (i.e. those having a CO 
strength of more than 5 $\kkms$ per SEST beam area) are numerous, but 
contribute only about a quarter to a third of the CO luminosity of the 
whole complex. This is similar to Galactic giant molecular clouds, where
e.g. Heyer et al. (1996) find most of the molecular mass to be in extended,
low column-density regions. Cloud 6 is embedded in a four times stronger 
radiation field (Table 1) and contains a very similar number of clumps per 
unit area. Two thirds of these are bright with {\it I(CO)} $>$ 5 $\kkms$ per 
SEST beam, but cloud 6 lacks the extended diffuse gas seen in cloud 35/36. 
In this complex, the contrast ratio is of order ten. Apparently, 
the fourfold increase in radiation density has resulted in the removal
of virtually all the low column density CO gas. As the high column density 
hydrogen gas will be practically unaffected, the value of {\it X} has increased 
in cloud 6 more or less commensurate with the increase in radiation density 
and CO removal. 

Cloud 32 experiences a radiation density a factor of three 
over that in cloud 6, i.e. over an order of magnitude more than 
that of Cloud 35/36. There is no trace of interclump gas. 
The number of clumps per unit area is still very similar, but
the fraction of {\it bright} clumps is only a third, down by more a factor 
of two from Cloud 6. Many of the weaker clumps are hardly discernible. 
We conclude that the further increase in radiation density in cloud 32 
is eroding even the dense CO clumps that are surviving reasonably well 
in cloud 6, resulting in a further decrease of the {\it I(CO)}/{\it N(H$_{2}$)}
ratio, i.e. a further increase in {\it X}.

It is of interest that the low-excitation clouds south of
N 159 have {\it X} values only a few times $\twcm \cm2$ rather similar to 
the canonical value of {\it X} in the Milky Way. In these clouds,
the lack of dissociating radiation apparently allows the CO to fill
most of the $\h2$ volume, nothwithstanding the lower CO abundance.

\subsubsection{Dependence on metallicity}
 
\begin{figure}
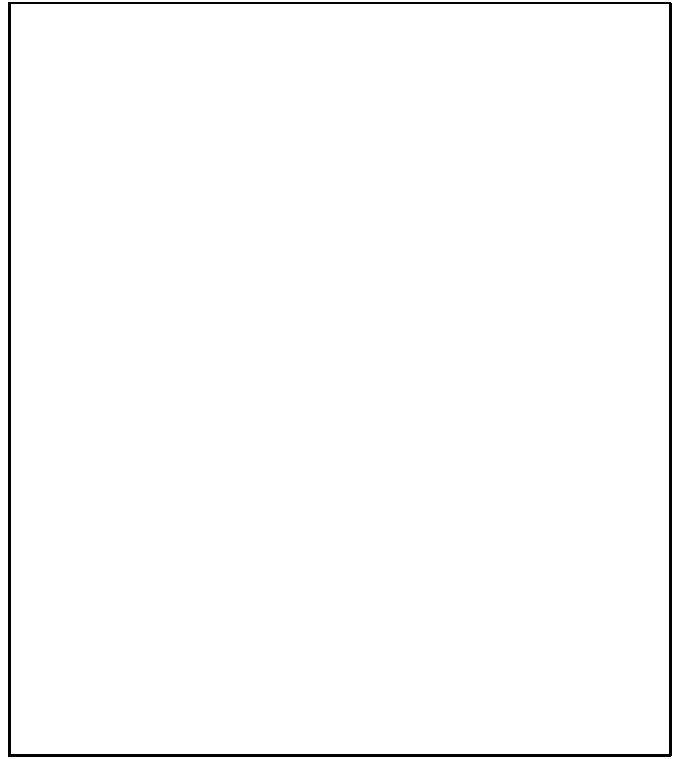
 
\picplace{10cm} 
\caption[]{Dependence of {\it X} on metallicity. a. At left: the ratio of 
{\it X} to $\sfir$/$\nh$ as a function metallicity [O]/[H]. The linear 
regression is drawn as a solid line. Only one line is drawn, as inclusion 
of the Milky Way points does not perceptibly change the result. b. At right: 
{\it X} as a function of metallicity [O]/[H] regardless of ambient radiation 
field. Global Milky Way points are indicated corresponding to `{\it Y}' from 
Bloemen at al.(1990 -- filled circle) and the more commonly used `{\it X}' 
from Bloemen et al. (1986 -- open circle). Regression lines are marked 
for the sample galaxies only, and for the sample galaxies with the addition
of the Galactic `{\it Y}' point (steeper line).
}            
\end{figure} 

The ratio of {\it X} to $\sfir$/$\nh$ for the objects in Tables 1, 2 and 3
can now be compared to the metallicities given in Table 5. We have 
included the Milky Way by using the {\it Y} and $\sfir$/$\nh$ values 
tabulated as R2 tot R5 by Bloemen et al. (1990 -- their Table 3) 
and the metallicities from Shaver et al. (1983). We have determined 
linear regressions with and without the Milky Way points, and for 
{\it X}--$\sfir$/$\nh$ dependences with exponents 0.9 and 1.0. We find: 

\begin{flushleft}
log {\it X} = 0.9$\pm$0.1 log($\frac{\sfir}{\nh}$)
      - 3.5$\pm$0.2 log$\frac{[O]}{[H]}$ + 34.6$\pm$2.2 \\
({\it n = 10; regression coefficient r$^{2}$ = 0.78}) \hfill (3a)\\
\end{flushleft}

Inclusion of the Milky Way points changes this to: 

\begin{flushleft}
log {\it X} = 0.9$\pm$0.1 log($\frac{\sfir}{\nh}$)
      - 3.2$\pm$0.1 log$\frac{[O]}{[H]}$ + 34.3$\pm$3.1 \\
({\it n = 14; r$^{2}$ = 0.85}) \hfill (3b)\\
\end{flushleft}

This is illustrated in Fig. 4a, which shows the dependence on [O]/[H]
for a linear (exponent 1.0) dependence of {\it X} on $\sfir$/$\nh$. As is 
clear from Fig. 4a, there is almost perfect agreement of the Milky Way points
with the relation determined for the sample galaxies alone. We have also 
empirically determined the dependence of {\it X} on metallicity, ignoring any 
dependence on radiation density. This yields (see also Fig. 4): 

\begin{flushleft}
log {\it X} = -2.7$\pm$0.3 log$\frac{[O]}{[H]}$ + 11.6$\pm$1.0 \\
({\it n = 10; r$^{2}$ = 0.90}) \hfill (4)\\
\end{flushleft}

Here, the nominal global Milky Way {\it Y}-point (Bloemen et al. 1990) is low 
compared to the relation defined by the sample galaxies, while the 
commonly used value {\it X} = 2.3 $\times$ 10$^{20}$ cm$^{-2} (\kkms)^{-1}$
provides a very good fit.

\begin{table}
\caption[]{Adopted Abundances} 
\begin{flushleft} 
\begin{tabular}{lcl} 
\hline\noalign{\smallskip}
Galaxy       & [O]/[H]      & References  	\\
             & 10$^{-4}$    &             	\\ 
             &              &             	\\
\noalign{\smallskip}
\hline
\noalign{\smallskip}
LMC	     & 2.60$\pm$0.40 & 1, 2, 3, 4      \\
SMC	     & 1.05$\pm$0.20 & 1, 2, 3, 4      \\
NGC 55       & 2.15$\pm$0.25 & 5, 6, 7         \\
NGC 1569     & 1.35$\pm$0.15 & 2, 5	       \\
NGC 4214     & 2.05$\pm$0.15 & 2, 7, 8         \\
NGC 4449     & 3.0$\pm$1.0   & 2, 9, 10	       \\
NGC 6822     & 1.60$\pm$0.15 & 2, 5, 11, 12, 13 \\
NGC 604      & 2.40$\pm$0.30 & 14, 15, 16      \\
NGC 595      & 2.40$\pm$0.30 & 14, 16          \\
NGC 5461     & 3.30$\pm$0.40 & 9, 17, 18       \\ 
\noalign{\smallskip}
\hline
\end{tabular}
\end{flushleft} 
References for abundances: 1. Dufour (1984) and references cited; 
2. Skillman et al. (1989a) and references cited; 3. Campbell (1992); 
4. Pagel et al. (1992); 5. Talent (1980); 6. Webster $\&$ Smith 1983; 
7. Stasinska et al. (1986); 8. Kobulnicky $\&$ Skillman 1996; 9. McCall 
(1982); 10. Hunter et al. (1982) 11. Skillman et al (1989b); 12. Kinman 
et al. (1979) 13. Pagel et al. (1980); 14. Kwitter $\&$ Aller (1981); 
15. Diaz et al. (1987); 16 Vilchez et al. (1988); 17. Rayo et al. (1982); 
18. Evans (1986). \\
\end{table} 

The dependence of {\it X} on [O]/[H] alone, ignoring $\sfir$/$\nh$ effects,
found here is significantly steeper than the result log {\it X} $\propto$
1.5 log [O]/[H] found by Sakamoto (1996) from modelling radiative transfer
and excitation of CO in clumpy molecular clouds. However, that result did 
not take into account the full effects of photodissociation of CO especially
on the interclump gas.

With respect to steep dependences on metallicity, we note that Garnett 
et al. (1995) have shown that [C]/[O] $\propto$ [O]/[H]$^{0.43\pm 0.09}$, 
so that [C]/[H] should be proportional roughly to [O]/[H]$^{1.5}$. If the 
CO abundance is solely determined by the carbon abundance, [CO]/[H] 
likewise will be proportional to [O]/[H]$^{1.5}$; if the oxygen 
abundance plays a role this may increase to [O]/[H]$^{2.5}$. The 
strength of the radiation field experienced by CO is proportional to 
the photon flux diluted by dust extinction. To first order, we may equate 
the decrease in dust shielding with the decrease in dust abundance. As 
the dust-to-gas ratio in galaxies is about proportional to [O]/[H]
(see e.g. Issa et al. 1990), we expect photodissociation alone to gain 
in importance as roughly [O]/[H]$^{-3}$ when metallicity 
decreases. Because the effects of photo-dissociation are highly
non-linear, and depend critically on the balance between ambient
radiation field and local column densities, a more quantitative 
estimate of the effect of metallicity can only be obtained by 
detailed modelling, which should treat photodissociation more 
rigorously and take into account the structure and column density 
distribution of the molecular clouds experiencing the dissociating 
radiation ({\it cf.} Maloney 1990a). This is beyond the scope of this paper.

\subsection{H$_{2}$ masses}

The results given in Tables 1 and 2 show the presence of significant 
amounts of molecular hydrogen in both the LMC and the SMC. At the 
(CO-selected) positions sampled, $\h2$ locally dominates the
interstellar gas. Table 3 suggests that this is also true in the
other galaxies. 

Can we extrapolate the results obtained so far to estimate the {\it total} 
amount of $\h2$ in the sample galaxies? Application of the mean {\it X} 
value from Table 1 to the LMC CO results obtained by Cohen et al. (1988) yield 
$M(H_{2})$ = 0.8 $\times$ 10$^{8}$ $M_{\odot}$. A more 
detailed treatment, applying the individual {\it N(H$_{2}$)} values from Table 
1 to the cloud complex sizes given by Cohen et al. (1988) and correcting 
for their CO sources not included in our Table 1, yields a higher value 
$M(H_{2})$ = 1.2 $\times$ 10$^{8}$ $M_{\odot}$. This result is  
very similar to that obtained by Cohen et al. (1988) (but see sect. 3.1). 
It corresponds to a global mass ratio of molecular-to-atomic hydrogen of 
0.2, much lower than the mass ratio of 1.9 found for the {\it individual 
CO clouds} in Table 1. It implies that in the LMC, about 12$\%$ of all the 
interstellar gas, including helium, is in molecular form.

We have also applied eqn. (2) to the total {\it far-infrared} emission of
the LMC, yielding a total hydrogen mass $M'_{\rm H}$ = 1.1 $\times$ 10$^{8}$ 
$M_{\odot}$, much less than the observed total HI mass {\it M(HI)} = 5.4 
$\times$ 10$^{8}$ $M_{\odot}$ (McGee $\&$ Milton 1966). Apparently, the LMC
contains a significant fraction of relatively cool dust not significantly
contributing to the total infrared luminosity. However, we may still
estimate the total amount of {\it $\h2$ associated with the warm dust}
by assuming the mean $\h2$/HI mass ratio from Table 1 to apply to all 
sources of warm $\h2$: we then find $M(H_{2})$ = 0.7 $\times$ 
10$^{8}$ $M_{\odot}$. This rougher method thus underestimates the actual
amount of $\h2$ by about 30$\%$.

Following the same procedures for the SMC, we find $M(\h2)$ = 0.50$\pm$0.05 
$\times$ 10$^{8}$ $M_{\odot}$, implying $M(H_{2})$/{\it M(HI)} = 0.1, or 
7$\%$ of all interstellar gas in molecular form. This is a lower limit, 
since the CO observations by Rubio et al. (1991) cover only part of the SMC. 
Indeed, extrapolation to the total infrared luminosity of the SMC yields 
$M'_{\rm H}$ = 1.75 $\times$ 10$^{8}$ $M_{\odot}$ which nevertheless still 
falls short of the total HI mass {\it M(HI)} = 5 $\times$ 10$^{8}$ 
$M_{\odot}$ (Hindman 1967). Applying the mean $\h2$/HI mass ratio from 
Table 2, we obtain $M(H_{2})$ = 1.0 $\times$ 10$^{8}$ $M_{\odot}$. 

\begin{table}
\caption[]{Global $\h2$ mass estimates} 
\begin{flushleft} 
\begin{tabular}{lcccccccccr} 
\hline\noalign{\smallskip}
Galaxy   & {\it M(HI)}   & $M'_{\rm H}$ & $M(\h2)$    & $\frac{M(\h2 )}{M(HI)}$ & $\frac{M(\h2 )}{M_{\rm gas}}$ \\
         & \multicolumn{3}{c}{10$^{8}$ $M_{\odot}$}   & 		       & \\
\noalign{\smallskip}
\hline
\noalign{\smallskip}
LMC	 &  5.4       & 1.1	     & 1.0	  & 0.19	   & 0.12 \\
SMC	 &  5.0       & 1.8	     & 0.75	  & 0.20	   & 0.12 \\
NGC 55   & 18.6	      & 4.8	     & 2.9	  & 0.16	   & 0.10 \\
NGC 1569 &  1.4	      & 0.65  	     & 0.5        & 0.35   	   & 0.20 \\
NGC 4214 & 11.2	      & 1.4  	     & 1.0        & 0.09           & 0.06 \\
NGC 4449 & 55	      & 21     	     & 8.7 	  & 0.16	   & 0.10 \\
NGC 6822 &  1.5	      & 0.5  	     & 0.4        & 0.27           & 0.16 \\
\noalign{\smallskip}
\hline
\noalign{\smallskip}
Mean	 &	      &		     &		  & 0.20  	   & 0.12 \\
	 &	      &		     &		  & $\pm$0.04	   & $\pm$0.02 \\
Milky Way & 48	      & ---	     & 12	  & 0.25	   & 0.15 \\
\noalign{\smallskip}
\hline
\end{tabular}
\end{flushleft} 
\end{table} 

The galaxies listed in Table 3 were not sampled extensively in CO, so that
we can only extrapolate from the total infrared luminosity. However,
the example of the LMC suggests that this extrapolation
is accurate to about 30$\%$. The results are given in Table 6, which 
also includes the global results for the Milky Way given by Bloemen et 
al. (1990). Table 6 shows that molecular hydrogen, although dominating 
the interstellar medium near star formation regions, occurs much less 
predominantly in irregular dwarf galaxies as a whole. Globally, the 
total mass of atomic hydrogen is typically five times higher than 
that of molecular hydrogen. On average, about one eigth of the interstellar 
gas mass is in molecular form. These fractions are surprisingly 
close to those of the Milky Way Galaxy as a whole, where the fraction of
molecular gas reaches a peak of 0.46 in the `molecular ring' at R = 4 -- 8 
kpc, and drops to 0.11 in the outer galaxy (Bloemen et al. 1990). If 
$X_{\rm Gal}$ is somewhat lower than assumed by Bloemen et al., as has 
been suggested by e.g. Bhat et al. (1986), the similarity of the Milky 
Way and irregular dwarf mean ratio is even more striking. Our result does 
not reproduce the apparent dependence of global molecular gas fraction 
on metallicity discussed by Tosi $\&$ D\'iaz (1990) and Vila-Costas $\&$ 
Edmunds (1992). We note that the latter expressed doubts on the physical 
significance of that result, and suggested that it might be an artifact of 
the CO to $\h2$ conversion used. Our result implies that this is indeed the 
case.

The molecular hydrogen fraction in this admittedly small sample appears
to be uncorrelated with metallicity, hence presumably dust-to-gas ratio.
This is somewhat surprising as $\h2$ molecules form on dust grain surfaces,
so that one would expect less $\h2$ in low metallicity environments poor
in dust grains. Our result may thus indicate that the formation of $\h2$
is so efficient, that it is to first order independent of the dust
abundance. Alternatively, it may reflect a selection effect. Higher
metallicity environments provide more shielding and therefore may have a 
larger fraction of cold dust/molecular hydrogen than lower metallicity 
environments. In the presence of warm dust, IRAS fluxes poorly sample 
cold dust, so that we may increasingly have underestimated the total 
amount of $\h2$ for the higher metallicity galaxies.

\section{Conclusions}

\begin{enumerate}

\item IRAS far-infrared surface brightnesses and HI 
column densities are used to indepently estimate $\h2$ column densities 
towards CO clouds observed in the LMC and SMC. Generally, in these clouds 
$\h2$ mass surface densities exceed those of HI by a factor of about 1.5 
on average. This is in contrast to the global $\h2$ to HI mass ratios
which are of the order of 20$\pm$10 $\%$.

\item By combining the newly derived $\h2$ column densities with
published CO intensities, the CO to $\h2$ conversion factors {\it X} are
determined to be $X_{\rm LMC}$ = 1.3$\pm$0.2 $\times$ 10$^{21}$ molecules
cm$^{-2}$ ($\kkms$)$^{-1}$ and $X_{\rm SMC}$ = 12$\pm$2 $\times$ 10$^{21}$ 
molecules cm$^{-2}$ ($\kkms$)$^{-1}$. The global mass of (warm) molecular
hydrogen is estimated to be $M(\h2)$ = 1.0$\pm$0.3 $\times$ 10$^{8}$ 
$M_{\odot}$ for both LMC and SMC.

\item On average somewhat higher molecular to atomic hydrogen mass surface
densities are found in the irregular dwarf galaxies NGC 55, NGC 1569, NGC
4214, NGC 4449 and NGC 6822, as well as in the extragalactic HII region 
complexes NGC 604, NGC 595, both in M 33, and NGC 5461 in M 101. The 
{\it X-}values derived for the HII regions and NGC 4449 are comparable 
to that of the LMC, while the {\it X}-values derived for NGC 55, NGC 4214 
and NGC 6822 are typically two to four times higher; NGC 1569 has 
a very high value comparable to that of the SMC.

\item  Analysis suggests that the CO to $\h2$ conversion factor {\it X}
is linearly dependent on the strength of the ambient radiation field
per nucleon, and inversely dependent on a steep function of metallicity 
[O]/[H]: log {\it X} = 0.9$\pm$0.1 log$\frac{\sfir}{\nh}$ - 3.5$\pm$0.2 
log$\frac{[O]}{[H]}$ + 34.6$\pm$2.2. If the dependence on radiation field
is neglected, the relation log {\it X} = -2.7$\pm$0.3 log$\frac{[O]}{[H]}$ + 
11.6$\pm$1.0 also fits the data. Similarly derived Milky Way values
fit these same relations. They are interpreted as the 
result of selective photodissociation of CO under conditions of high
radiation field energy densities and poor shielding and selfshielding
in low-metallicity environments. Thus, over the parameter range
studied, the CO content of galaxies varies strongly as a function
of conditions.

\item Estimates of the global (warm) $\h2$ to HI mass ratios and
the (warm) $\h2$ gas fractions yield very similar results for all
galaxies. On average, $M(\h2)$ = 0.20 {\it M(HI)}, and $M(\h2)$ = 0.12
$M_{\rm gas}$. These ratios are very close to the global Milky Way 
ratios: the global warm $\h2$ fraction in irregular dwarf galaxies
appears to be very similar to that of our Galaxy, notwithstanding
the large differences in total mass, luminosity, metallicity and 
observed CO luminosity.

\end{enumerate}

\acknowledgements 

It is a pleasure to thank J. Kamphuis, J. Stil and F. Sloff for making 
their results available in advance of publication, and J.B.G.M. Bloemen 
for drawing attention to the Galactic results.

\end{document}